\documentclass[11pt]{article}
\usepackage{graphicx,color}
\usepackage{epstopdf}
\begin{document}
\title{Solution to the gauge-Higgs analyticity paradox }
\author{Michael Grady\\
Department of Physics\\ State University of New York at Fredonia\\
Fredonia NY 14063 USA\\ph:(716)673-4624, fax:(716)673-3347, email: grady@fredonia.edu}
\date{\today}
\maketitle
\thispagestyle{empty}
\begin{abstract}
The Fradkin-Shenker theorem proves analyticity in a region that connects Higgs to confinement regimes, precluding a 
phase transition. This conflicts with a simpler analyticity argument applicable to any symmetry-breaking phase transition 
that requires the phase diagram to be bifurcated.  A flaw in the Fradkin-Shenker and related Osterwalder-Seiler proofs is 
found which removes this paradox. Higgs and Confinement regions are everywhere separated by a phase boundary. A new order 
parameter allowing this transition to be traced with Monte-Carlo simulations without gauge fixing is introduced.
\end{abstract}


\renewcommand{\baselinestretch}{1.5}
One of the few rigorous results in lattice gauge theory is the Fradkin Shenker (FS) theorem\cite{fs}.  
It states that when both gauge and Higgs fields are in the fundamental representation
a finite-width region exists in the gauge-Higgs coupling plane
where 
expectation values of local operators are analytic functions
of the couplings.  Consequently, the Higgs and confinement phases are continuously connected,
not separated by a phase transition, despite differing qualitatively.
That both phases are massive makes this conceivable. 
The FS theorem applies to the fixed-magnitude Higgs
field. It specializes
an earlier theorem by Osterwalder and Seiler\cite{os} (OS) which proved a region of analyticity
in the Higgs phase for a variable-magnitude Higgs, and for pure gauge theory at strong coupling.
The proof relies on a 
convergent cluster expansion, a technique developed earlier for $p(\phi )_2$ theories
by Glimm, Jaffee, and Spencer(GLS)\cite{gls}.  Although the series converges, it is shown below that
individual terms being summed can themselves be non-analytic, which invalidates the proof. In fact a simpler
analyticity argument based on the exactness of the broken symmetry leads to the opposite conclusion, that 
Higgs and confinement regions are everywhere separated by a phase transition.
   
The system has action
\begin{equation}
S=-\beta \sum _p \frac{1}{D}(\mbox{tr} (U_p ) -D)-\lambda \sum _{\vec{r},\mu } 
\frac{1}{D}\mbox{tr} (\phi ^{\dagger}(\vec{r})U_\mu (\vec{r})\phi (\vec{r}+\hat{\mu}))\label{eqn1}
\end{equation}
where $U_\mu $ are gauge fields on links  belonging to the fundamental
representation of
a gauge group, $U_p$ is the elementary plaquette made from
the product of four links $UUU^{\dagger}U^{\dagger}$, and $\phi$ are site-based
matrix-valued Higgs fields also in the fundamental representation.  $\beta \propto 1/g^2$ is the
inverse gauge coupling and $\lambda$ is the 
Higgs coupling. $D$ is the representation dimension. For complex representations, real parts
are taken.
The cases of 3-d Z2, which is self-dual, and 4-d SU(2) are emphasized here.
Fig.~1 shows the phase diagram for Z2.
The analyticity region(AR) is the striped area and is
representative - not drawn to scale.
All theories in the above category have similar analyticity regions and a Higgs 
transition that presumably ends before reaching it.  Monte Carlo simulations show
a generally strong transition at high $\beta$ becoming weaker as $\beta$ is lowered and
eventually looking like a crossover. The exact terminus of the transition
(critical point) in the different theories is somewhat disputed, as is whether it is first
or second order at the endpoint
(orders are group dependent)\cite{ghgen,z2}.  Another transition 
joining the Higgs 
transition originates on the
pure-gauge axis in some theories, separating Coulomb and confinement phases.

\begin{figure}[ht]
                      \includegraphics[width=4in]{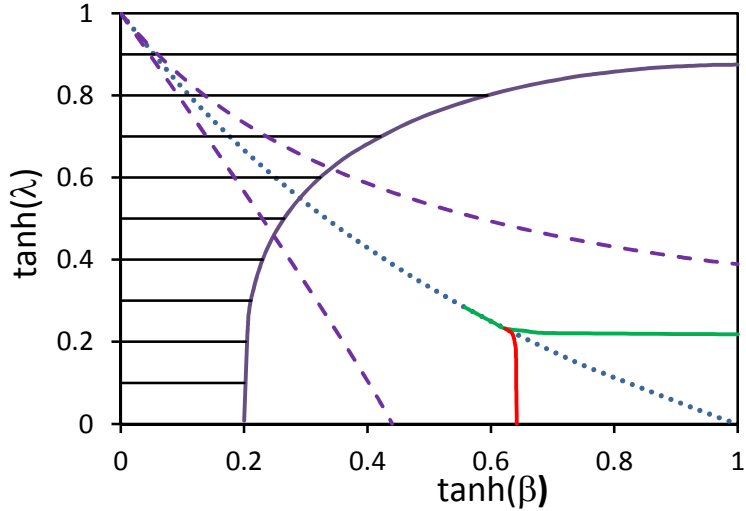}
                                  \caption{Gauge-Higgs phase diagram for the 3-d Z2 theory.  Phase transitions (on right) are solid,
                                  based on ref. \cite{z2}.
                                  Self-dual line is dotted.  A representative Fradkin-Shenker analyticity region(AR) is striped.  
				Dashed lines represent a possible alternative AR based on roughening transitions, with on-axis
				locations from ref. \cite{rougheningising}.}
          \label{fig1}
       \end{figure}

At $\beta = \infty$ is a spin model,  3-d Ising in the 3-d Z2 case and
O(4) Heisenberg for 4-d SU(2). These have well known symmetry-breaking magnetic phase transitions.
The phase transition survives entering
the phase diagram, in some cases changing order.
The symmetry-breaking nature of the phase transition
becomes hidden by the prohibition of local symmetry breaking(Elitzur's theorem\cite{elitzur}).
The same issue confronts the Higgs mechanism in the standard model, which no one doubts
is driven by spontaneous symmetry breaking.  Symmetry breaking becomes visible in
Landau gauge, which leaves a remnant {\em global} symmetry unfixed.  The remnant symmetry 
is the symmetry broken in the Higgs
phase, driving the Higgs mechanism. The symmetry-breaking nature
of the Higgs transition conflicts with
the FS theorem.  A phase transition line separating symmetry-realized and symmetry-broken
regions cannot end in a critical point, as in Fig.~1, because an analytic
function which vanishes in a finite region vanishes everywhere.  For an exact symmetry
this applies to the order parameter in the symmetry-realized phase. Everywhere it touches the symmetry-broken region is
non-analytic. Thus
the transition must continue until it hits an edge of the phase diagram. 
There is only one option for the endpoint (explained in detail later), the upper left corner.
Thus a paradox exists. The FS theorem claims analyticity
in the same region that the simple argument above, also based on analytic function theory, predicts a phase transition.  
Numerical 
observation of apparently continuous crossovers in energy quantities in this
region cannot be trusted to disprove the existence of a phase transition, because some
phase transitions are only weakly displayed in energy quantities. 
Generally, using numerical methods
applied to energy quantities alone, one can detect only
phase transitions for which the infinite-lattice specific heat becomes infinite.
A specific heat peak is observed growing with lattice size if
the critical exponent $\alpha >0$. However, if $\alpha <0$ the specific heat remains 
finite at the critical point and shows little finite-size effect.  In this case it is a derivative of
the specific heat which diverges. It may develop a cusp or a curvature cusp,
for instance,  which are difficult to distinguish numerically from continuous crossovers.

If one has an order parameter to study, however, then weak phase transitions are detectable from
finite-size crossings of Binder cumulant or normalized correlation length. Both
have opposite scaling behavior on opposite sides of a phase transition and are lattice-size invariant
at the phase transition.  Caudy and Greensite(CG) have studied 
the Higgs transition in 4-d SU(2) using Landau gauge\cite{cg}.  A regular gauge-invariant Monte Carlo simulation 
is run and gauge configurations are
transformed to lattice Landau gauge which seeks to
maximize tr$(U)$ for all links.  Since a global gauge transformation 
leaves tr$(U)$ invariant, that remains an exact symmetry subject to
spontaneous symmetry breaking.  To the extent that gauge fixing can produce many links close to the identity, the
system approaches the spin model (where links are unity).  It resembles
a spin glass
where some of the interactions are not fully ferromagnetic.  As $\beta$ decreases interaction
disorder increases until eventually ferromagnetism is lost.
It can be regained by increasing $\lambda$ which favors order.  This explains
the slope of the phase transition line. 
The Higgs field expectation value $<\! \phi \! >$ is the order parameter.
CG find a definite symmetry breaking for all $\beta $ tested. To rationalize
this result with the FS theorem, they have suggested the phase transition might at some point become non-thermal,
and turn into something like a Kert\'{e}sz line\cite{kertesz}. However a Kert\'{e}sz line occurs when {\em explicit} 
symmetry breaking smooths
a singularity, such as an Ising model in a weak magnetic field.  The percolation transition persists, but is not associated
with an energy singularity. But neither is there an order parameter singularity in this case.  What is needed
for the gauge-Higgs model is for the energy to be non-singular even though the order parameter is singular, in other words
for the energy to simply not care about the order parameter anymore. Such a critical point would be unique in the 
annals of statistical mechanics.
If the fields carry energy
then how could a singularity in the field not affect the energy in some way?  If the energy no
longer cared about the field, then it would randomize,  
preventing magnetization.  A system magnetizes due to the dependence of the free energy on magnetization. 
This requires the free energy and magnetization to be correlated, which will transmit singularities in the 
magnetization to the free energy, because these are both sharp quantities in the thermodynamic limit.

CG give another argument to discount the observed phase transition.
A similar study was done in the lattice Coulomb gauge, which maximizes the traces of links 
in only three of four directions, leaving a much larger remnant gauge symmetry, global in three directions
and local in one.  
Expectation values of gauge links pointing in the fourth direction $<\sum U_ 4 >$ summed over each 
hyperlayer are the order parameters.  They find symmetry breaking in this quantity also, however in one region
it is along a line above the Landau gauge result. They argue
this apparent ``gauge ambiguity" could signify an artifact. 
However, there is a good reason why the Coulomb gauge result may differ from Landau
gauge.  These are different order parameters testing different symmetries (see Appendix). 
There may well be two parallel phase transitions in part of the phase diagram,
with an interesting phase between them, in which the 
Higgs symmetry is spontaneously broken but the gauge fields are still confined.  Without a long range force to
drive the Higgs mechanism a massless Goldstone boson may exist here. Another possible 
explanation for the Coulomb
gauge result differing are potential problems with Gribov copies,
local gauge condition minima. So
this result should probably be checked for possible systematic errors from gauge fixing,
and a finite size scaling study be undertaken to precisely fix the location,
before a separate transition is definitively concluded.  
To demonstrate that gauge fixing isn't somehow creating a false signal, below
I introduce a new order parameter for the Higgs transition that doesn't require gauge fixing.
The phase transition it sees corresponds to that seen using Landau gauge, which makes sense since both are sensitive to 
the Higgs symmetry breaking.  So at least this transition is not gauge-dependent. 

Consider the shape of the analyticity region (AR). FS have computed this explicitly for the Z2
case shown.  Based on inequalities, it is a lower limit of the true AR.  
The strong coupling
expansion in the pure gauge theory appears to converge up to the roughening transition which for pure-gauge SU(2) is 
around $\beta=1.9$
\cite{roughening}.  The first actual singularity appears to occur here.  
Roughening also exists for the 3-d Ising model and its dual, the 3-d Ising gauge theory\cite{rougheningising}.
These transitions presumably enter the phase diagram.  Since increasing $\lambda$ increases order, one would expect
that to stay on this transition one would have to compensate 
by lowering $\beta$.  This results in the ``true" AR (roughening transition line) sloping oppositely to the FS AR.  
For 3-d  Z2, a dual
roughening transition exists in the Higgs phase. 
Opposite slope of the FS AR from the likely behavior of the physical AR is not necessarily inconsistent,
because the roughening transitions could conceivably join the main transition at some point.  However,
the opposite slope of the FS AR from an order-disorder contour still seems odd.

The upper left corner of the phase diagram has a demonstrable singularity\cite{me}. 
The $\beta =0$
theory is solvable.  In axial gauge, no interactions perpendicular
to the gauge-fixed direction remain, resulting in a set of 1-d Ising or Heisenberg models. 
These are
disordered for all finite $\lambda$ but are ordered at $\lambda =\infty$.  The 1-d Ising model
has a phase transition at $T=0$, accompanied by an essential singularity, in our notation at  $1/\lambda =0$.
This result can also be obtained by a single integration in unitary gauge. The Heisenberg case is
somewhat less singular but still the correlation length diverges\cite{heisenbergsingularity}.  Thus a singularity
exists deep within the FS AR.  This makes a natural endpoint for the phase transition and is the self-dual line
endpoint. However, it is possible this is an isolated singular
point. The FS theorem is evaded simply because the first term in the cluster expansion is itself singular
here. 

Summarizing, two different analyticity arguments, the FS theorem and the exact symmetry-breaking
argument, have opposite conclusions. One forbids a phase transition whereas the other requires one.  Accepting FS means
accepting the concept of an unprecedented exact-symmetry non-thermal transition. The other possibility is a flaw in the FS argument, 
with the AR region sloping in the physically sensible
direction, pinching down at the upper left corner, allowing the phase transition through (alternative AR in Fig.~1). 
The FS theorem relies on proving
the convergence of the cluster expansion. It uses unitary gauge which removes the Higgs field entirely.
The Higgs action becomes
\begin{equation}
S_H =-\lambda \sum _{\vec{r},\mu} \frac{1}{D} \mbox{tr}U_\mu (\vec{r}) .
\end{equation}
Consider the expectation value of a local operator $\cal{F}$ such as the average plaquette.  The cluster expansion rewrites
the individual plaquette 
Boltzman factor 
\begin{equation}
\exp (-\beta S_{p}) = 1+p_{p}
\end{equation}
For small $\beta$, $p_p$ are small. The Higgs  action, now local on links,
is absorbed into the link measure $dU$.  The cluster expansion for $<\cal{F}>$ is given by
\begin{equation}
 <\mathcal{F}>= \sum_{Q_{1}(Q_{0})}\int dU \mathcal{F}\prod _{p\in Q_{1}}p_{p}\frac{Z(\overline{Q_{1} \cup Q_{0}})}{Z} .
\end{equation}
Here $Q_0$ is the set of plaquettes connected to $\cal{F}$. The expansion is over all connected sets of plaquettes $Q_1$
connected with $Q_0$. $Z$ is the partition function and $Z(\overline{Q_1 \cup Q_0})$ is the partition function
missing all plaquettes in $Q_1 \cup Q_0$ and any touching its boundary. For details see FS\cite{fs} and OS\cite{os}.
As $Q_1$ grows terms have a larger number of the small factors $p_{p}$ which for small $\beta$ form
a convergent series.  The ratio of the two partition functions sums the disconnected diagrams and does not
spoil the series convergence for small enough $\beta $ or large enough $\lambda $. Small $\beta$ and large $\lambda$ both 
aid convergence
which explains the slope of the FS AR.  Quoting FS, ``Analyticity of $<\cal{F}>$ in (the couplings) (follows),
because the series converges uniformly and the terms are each analytic."  There seems little doubt the series 
converges in the region claimed.
However, the second condition that the individual terms are analytic is not addressed in either FS or OS. Presumably it
was thought too obvious to require proof. Recall that non-analyticity of an individual term is the loophole that
allows a singularity at the upper corner. Could this problem be more widespread? The suspicious factor is the 
ratio of the two partition functions, one missing some of its plaquettes. Since a partition function is the sum of an infinite
number of terms in the thermodynamic limit, here we have a finite ratio of two infinite quantities. It is precisely
such a ratio that gives rise to thermodynamic singularities. It is not at all apparent that such a factor is 
singularity free.  Consider the simplest case of $Z$(missing a single plaquette)$/Z$ which can be written $<\exp (\beta S_p)>$ 
where $S_p$ is the single plaquette action.  If we expand the exponential this contains a term of $<S_p >$ the expectation 
value of the average
plaquette itself.  So in the cluster expansion for the average plaquette, some of the terms on the RHS
{\em also} include factors of the average plaquette and other more complex expectation values.  
Thus if the average plaquette has
a singularity, there are singularities on both sides of the equation which seems perfectly consistent regardless of
convergence. Therefore a convergent cluster expansion is not sufficient to prove analyticity. It is consistent
with expectation values either being singular or not.  One must also 
prove that each ratio of partition functions on the RHS is itself analytic. This flaw is common to both FS and the second part of 
OS which concerns the Higgs mechanism.  It may be possible to save the first part of OS which covers
the pure gauge theory,
by {\em first} proving exponential clustering from a symmetry argument.  Singularities in expectation values arise
from massless excitations and infinite-range forces. Say the plaquette-plaquette correlation function follows a power law.
Then the $n$th derivative of the average plaquette with respect to $\beta$ is an integrated $n+1$-point function which
diverges for large enough $n$.  This will not happen for a massive 
theory where correlation functions fall exponentially (exponential clustering). So if one can prove exponential clustering
independently of the cluster expansion, that may serve as input to an analyticity proof.
In GJS, exponential clustering is proven early on 
from the $\phi \rightarrow -\phi$ symmetry of even-power $p(\phi)$ theories.
A similar symmetry is present in the pure gauge theory, $U \rightarrow -U$, but not for
$\lambda \neq 0$. 

The above doubt cast on the FS proof removes the paradox.  The fact that the symmetry being broken is exact yields a 
powerful argument in favor of a phase transition bifurcating the diagram.  It is not necessary to argue
away the transitions observed in Monte Carlo simulations.  
Higgs and confinement may both be massive phases but apparently they are different massive phases.

\section*{Appendix}

Here the symmetries of the gauge-Higgs system are examined more carefully and a new 
gauge-invaraint order parameter for the Higgs transition is
introduced. This helps ally any concern that the observed phase transitions could be
gauge-fixing artifacts.  The original action in eqn. \ref{eqn1} is invariant under the local gauge transformation, $V(\vec{r})$
\begin{equation}
U_{\mu}(\vec{r}) \rightarrow V(\vec{r})U_{\mu}(\vec{r})V^{\dagger}(\vec{r}+\hat{\mu}) ,\;
  \phi (\vec{r}) \rightarrow V(\vec{r}) \phi (\vec{r})
\end{equation}
There is also a separate global symmetry transformation, $W$, that affects 
only the $\phi$ fields,
\begin{equation}
\phi (\vec{r}) \rightarrow \phi (\vec{r}) W  .
\end{equation}
In the Landau gauge, a single
global $V$ remains and the $\phi$ transforms 
according to $\phi \rightarrow V\phi W$.  If $\phi$ takes an
expectation value, both the $W$ and $V$ symmetries are spontaneously broken. For the SU(2) case this gives an 
SU(2)$\times$SU(2) = O(4) space of broken vacua.  
To further separate these and probe the W symmetry alone, one can borrow a trick from the study of spin glasses, the 
``two real replica'' technique\cite{2rr}.  For a given gauge configuration (which has an associated 
Higgs configuration), one equilibrates another instance of the Higgs field $\phi _r$. 
Because this is not a genuine
 second Higgs field that the gauge
field ``knows'' about through detailed balance, a full Monte-Carlo equilibration 
is required.  Measured quantities approach asymptotic values exponentially with equilibration time, 
which can be adjusted to drive systematic errors as low as desired. Generally $e$-folding times are in the 
hundreds of sweeps and practical equilibration times in the low thousands.  Now there are two independently 
generated $\phi$ fields, each of which is from either a spontaneously broken ensemble of the $W$ symmetry or from an 
unbroken one.  If broken, the $\phi _r$ and $\phi$ will choose different symmetry-breaking directions. Indeed 
$\phi _r$ has a separate global symmetry $W_r$.  The order parameter is $<m>$, where
\begin{equation}
m=|\sum _{\vec{r}}\phi ^{\dagger} _{r}(\vec{r}) \phi (\vec{r}) | .
\end{equation}
The sum is over 4-space. The norm is O(4). Because both $\phi$'s 
are equilibrated to the same gauge background they similarly adjust to the gauge topography.
The order parameter is 
gauge invariant, but sensitive to the W-symmetry.  For unbroken W-symmetry both $\phi $ and $\phi _r$ 
will have tunnelings within each configuration  which will destroy any overall correlation when spatially summed. 
However, for broken $W$ symmetry an O(4) set of spontaneous broken vacua arise, with
nonzero $<m>$.  Such an order parameter is usually used to find spin-glass order but it will also detect 
ferromagnetic order, which is the case here because the transition observed is consistent with the location
of the Landau-gauge transition seen by CG.  Fig.~2 shows the Binder cumulant,
$U_B = 1-<m^4>/(3<m^2 >^2 )$ for $\beta = 1.2$ and
various $\lambda$ on $16^4$ and $24^4$ lattices.  An equilibration study was first performed to ensure systematic errors
were less than 10\% of statistical errors. This required 1800 sweeps for the $24^4$ case.
One sees a crossing at $\lambda = 1.36$ (verified in upper region at 15 standard deviations and also in the normalized
correlation length). The susceptibility shows a
growing peak.  Preliminary fits to finite size scaling give a critical exponent 
$\nu =0.8$.  The current uncertainty is roughly $\pm 0.15$. 
This study is being extended to include more lattice sizes and better statistics to further narrow this
estimate and establish a connection with energy quantities.  The $\nu$ estimate predicts 
$\alpha = 2-d\nu = -1.2$ with about a 50\% uncertainty, the negative value consistent with a finite
specific heat.  Crossings are also observed at 
$\beta = 0.5$. The correlation between this order parameter and internal energy
is non-zero in the broken phase, which demonstrates that the energy ``cares'' about this symmetry breaking.  
Details will be reported elsewhere.

\begin{figure}[ht]
                      \includegraphics[width=4in]{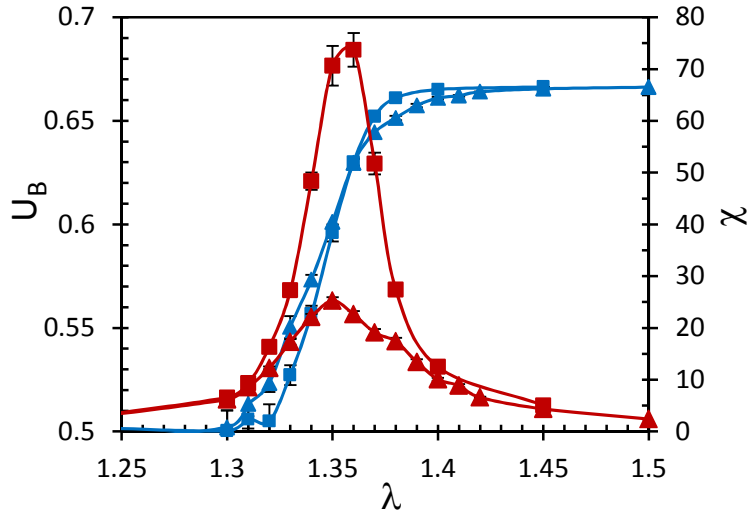}
                                  \caption{The Binder cumulant for the replica-field correlation order parameter (s-shape, left scale) 
and susceptibility(right scale), for  $\beta = 1.2$ on $16^4$ (triangles) and $24^4$ (boxes) lattices. Error bars computed from binned
fluctuations.}
          \label{fig2}
       \end{figure}

\end{document}